\documentclass[12pt,preprint]{aastex}

\usepackage[]{natbib}
\usepackage{graphicx}
\usepackage{amssymb}
\usepackage{color}
\usepackage{ulem}

\slugcomment{ver. 2.1}

\newcommand{\SUZAKU}{{\it Suzaku}}

\newcommand{\XMM}{{\it XMM-Newton}}

\newcommand{\CHANDRA}{{\it Chandra}}

\newcommand{\Suzaku}{{\it Suzaku}}

\newcommand{\UNITFLUX}{{\rm ergs~cm$^{-2}$~s$^{-1}$}}

\newcommand{\UNITCPS}{{\rm counts~s$^{-1}$}}

\newcommand{\UNITLUMI}{{\rm ergs~s$^{-1}$}}

\newcommand{\UNITNH}{{\rm cm$^{-2}$}}

\newcommand{\UNITSOLARMASSYEAR}{{\it M$_{\odot}$} yr$^{-1}$}

\newcommand{\ARCMIN}{{$'$}}

\newcommand{\ARCSEC}{{$''$}}

\newcommand{\NH}{{\it N$_{\rm H}$}}

\newcommand{\LX}{{\it L$_{\rm X}$}}

\newcommand{\KT}{{\it kT}}

\shorttitle{\SUZAKU\ Observation of an Erupting Young Stellar Object}
\shortauthors{Hamaguchi et al.}

\begin{document}

\title{\SUZAKU\ Observation of Strong Fluorescent Iron Line Emission from 
the Young Stellar Object V1647~Ori during Its New X-ray Outburst}

\author{Kenji Hamaguchi\altaffilmark{1,2}, Nicolas Grosso\altaffilmark{3,4}, Joel H. Kastner\altaffilmark{5}, 
David A. Weintraub\altaffilmark{6}, Michael Richmond\altaffilmark{5}}

\altaffiltext{1}{CRESST and X-ray Astrophysics Laboratory NASA/GSFC,
Greenbelt, MD 20771; Kenji.Hamaguchi@nasa.gov.}
\altaffiltext{2}{Department of Physics, University of Maryland, Baltimore County, 
1000 Hilltop Circle, Baltimore, MD 21250.}
\altaffiltext{3}{Universit\'e de Strasbourg, Observatoire Astronomique de
Strasbourg, 11 rue de l'universit\'e, 67000 Strasbourg, France}
\altaffiltext{4}{CNRS, UMR 7550, 11 rue de l'universit\'e, 67000 Strasbourg, France}
\altaffiltext{5}{Rochester Institute of Technology, 54 Lomb Memorial Drive, Rochester, NY 14623.}
\altaffiltext{6}{Vanderbilt University, Nashville, TN 37235.}

\begin{abstract}
The \SUZAKU\ X-ray satellite observed the young stellar object
V1647~Ori on 2008 October 8 during the new mass accretion outburst reported in August 2008.
During the 87~ksec observation with a net exposure of 40 ks, 
V1647 Ori showed a high level of X-ray emission with a gradual decrease in flux by a factor of 5 and then
displayed an abrupt flux increase by an order of magnitude. 
Such enhanced X-ray variability was also seen in \XMM\ observations in 2004 and 2005 during the 2003$-$2005 outburst, 
but has rarely been observed for other young stellar objects.
The spectrum clearly displays emission from Helium-like iron, which is a signature of hot plasma (\KT~$\sim$5 keV).
It also shows a fluorescent iron K$\alpha$ line with a remarkably large equivalent width of $\sim$600~eV.
Such a large equivalent width suggests that a part of the incident X-ray emission that 
irradiates the circumstellar material and/or the stellar surface is hidden from our line of sight.
\XMM\ spectra during the 2003$-$2005 outburst did not show a strong fluorescent iron K$\alpha$ line,
so that the structure of the circumstellar gas very close to the stellar core
that absorbs and re-emits X-ray emission from the central object
may have changed in between 2005 and 2008.
This phenomenon may be related to changes in the infrared morphology of McNeil's nebula between 2004 and 2008.
\end{abstract}
\keywords{stars: formation --- stars: individual (V1647 Ori) --- stars: pre-main sequence --- X-rays: stars}

\section{Introduction}

Certain young stars dramatically increase their mass accretion rates by orders of magnitude 
(from 10$^{-7}$\UNITSOLARMASSYEAR\ to 10$^{-4}$\UNITSOLARMASSYEAR).
These events are possibly triggered by thermal disk instabilities and are traced by dramatic increases in optical/IR luminosities;
Such eruptive pre-main sequence (PMS) stars are crudely classified as either FU Ori (FUor) or EX Lupi (EXor) types; 
the former are characterized by outbursts lasting decades, whereas the latter generally display shorter outbursts, of duration 
a few months or years \citep[see reviews in][]{Hartmann1996}.
Although PMS stars are known to be luminous X-ray sources,
FUor or EXor events have been reported only rarely, such that the level of PMS
X-ray activity during an outburst is poorly established.

The young stellar object (YSO) V1647 Ori, which is deeply embedded in the L1630 dark cloud 
($d \sim$400~pc, \citealt{Anthony-Twarog1982}), underwent a strong optical/NIR outburst in 2003 December.
This eruption afforded the first opportunity to record the sustained X-ray outburst of a rapidly
accreting YSO \citep{Kastner2004,Grosso2005,Grosso2006,Kastner2006}.
Multiple \CHANDRA\ and \XMM\ observations through this
outburst demonstrated that the average X-ray flux varied in the same way and on
approximately the same timescale as the optical and IR brightnesses, although the X-ray flux also varied strongly
(by a factor of up to 20) on timescales of less than a day.
During outburst,
the V1647~Ori X-ray spectrum had the characteristics of deeply embedded hot plasma 
(\NH\ $\sim$4.1$\times$10$^{22}$~\UNITNH, \KT\ $\sim$4.2~keV).
This result appears to be explained best as star-disk magnetic reconnection activity generated in association with
the episode of very rapid mass infall \citep{Kastner2004,Kastner2006}.

Following the V1647~Ori event,
the EXor candidate V1118~Ori recently displayed a mass accretion outburst;
followup X-ray observations of this object
detected a moderate enhancement in the X-ray flux that was correlated
with the optical/IR flux
\citep{Audard2005,Audard2009}.
\citet{Skinner2006,Skinner2009} reported that two FUor systems with ongoing historical outbursts, FU Ori and V1735 Cyg,
showed very hard spectra of \KT\ $\gtrsim$ 5~keV plasmas or equivalent despite relatively steady X-ray fluxes.
Systematic surveys of low-mass PMS (T Tauri) stars in the Orion nebula and the Taurus dark cloud
\citep[e.g.,][]{Preibisch2005b,Telleschi2007} 
indicate that accreting T Tauri stars are less X-ray active (by a factor of $\sim$2$-$3 on average) 
than non-accreting T Tauri stars, 
suggesting that mass accretion activity may in fact suppress X-ray activity.
Hence, the relationship between PMS mass accretion and high energy radiation remains unclear.

V1647~Ori began a new optical/NIR outburst at a certain time between 2008 Jan.\ 2 and Aug.\ 26
\citep{Itagaki2008,Aspin2008,Aspin2009}.
Our team later triggered an anticipated Target of Opportunity (ToO) observation of V1647~Ori with \CHANDRA\ 
on September 18 for 20~ksec,
in which the X--ray count rate was double the level observed by \CHANDRA\ on 2004 March 7
\citep{Kastner2004} during the previous outburst in 2003$-$2005 \citetext{Weintraub et al. in preparation}.
This new X-ray eruption offers us the first (and possibly unique) opportunity to measure X-ray emission 
during two outbursts from the same YSO.
We therefore proposed a ToO observation with the \SUZAKU\ X-ray observatory \citep{Mitsuda2007},
so as to obtain detailed X-ray spectral diagnostics and a $\sim$day timescale light curve of V1647~Ori
during its latest accretion outburst.
With this information,
we can derive the hottest plasma temperature and the intrinsic activity variation of V1647 Ori, unaffected by circumstellar
absorption; we can also constrain models of the plasma geometry and of
the response of neutral disk material to X-ray irradiation.
This paper focuses on the results of the Suzaku observation.

\section{Observation}

\SUZAKU\ observed V1647~Ori for $\sim$87~ksec on 2008 October 8,
when the YSO's optical/NIR fluxes leveled off at a brightness similar to that observed in 2004 Feb.$-$Mar.\ \citep{Ojha2008,Aspin2009b}.
At the time of this observation,
the observatory operated with three sets of telescope and detector systems ---
the thin-foil X-Ray Telescope (XRT) module \citep{Serlemitsos2007} with the X-ray CCD detector 
\citep[XIS: X-ray Imaging Spectrometer,][]{Koyama2007} on its focal plane,
and the Hard X-ray Detector \citep[HXD: ][]{Takahashi2007,Kokubun2007} with narrow collimator arrays.
The XRT+XIS system focuses X-rays with a half-power diameter (HPD) of $\sim$2\ARCMIN\ and covers 
the X-ray energy range up to $\sim$10~keV.
The XIS1 camera employs back-illuminated (BI) CCDs for the soft band sensitivity down to $\sim$0.3~keV,
while the other XIS cameras (XIS0, XIS3) use front-side illuminated (FI) CCDs with sensitivity down to $\sim$0.4~keV.
The HXD detects hard X-rays above 15~keV.

We started analysis from the cleaned event data in the distribution package,
pre-processed with the version 2.2.11.22.
The cleaned data excluded events recorded during high background periods,
mainly when the satellite passed the South Atlantic Anomaly (SAA) 9 times during this observation,
and earth occultations, each of which happened for $\sim$2.1~ksec in every \SUZAKU\ orbit ($\sim$5.4~ksec).
The data did not suffer telemetry saturation.
We thus selected events in good time intervals of all the XIS0, 1 and 3 sensors.
After this selection, the net exposure was 40,446~sec, i.e., $\sim$46\% of the observing time.
For the XIS {\tt arf} generation, we used the CALDB version released on 2009-02-03,
to better reproduce the recent growth of contamination on the XISs, in particular XIS0.
The FI (XIS~0+3) and BI sensors collected 1216 and 624 counts between 1$-$8~keV,
which include particle background and contamination from nearby sources by $\sim$26\% and $\sim$39\%, respectively.

We generated time-averaged HXD spectra,
following the standard analysis procedure\footnote{http://heasarc.gsfc.nasa.gov/docs/suzaku/analysis/abc/}
and using the version 2.0 tuned background\footnote{ftp://legacy.gsfc.nasa.gov/suzaku/data/background/pinnxb\_ver2.0\_tuned/}.
The spectrum was consistent with the typical cosmic X-ray background spectrum and showed no hint of signal from V1647 Ori.

\section{Image Analysis}

\begin{figure}[t]
\plotone{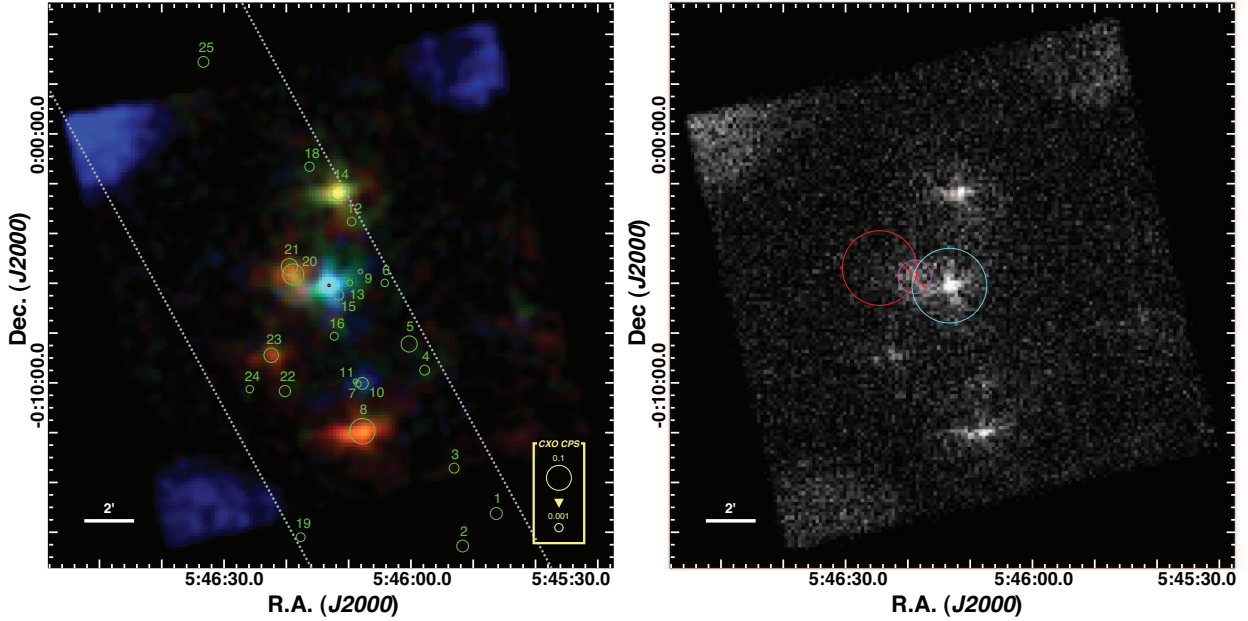}
\caption{
{\it Left}: XIS~0+1+3 color image of the V1647~Ori field 
(red: 0.5$-$2~keV, green: 2$-$4~keV, blue: 4$-$10~keV)
after correcting the absolute coordinates using a \CHANDRA\ result \citep{Simon2004}.
The image was smoothed by the Gaussian function with $\sigma$ =3~image pixels
($\sim$25\ARCSEC)
and displayed in a linear scale for the 5~counts pixel$^{-1}$ range in each color,
offset at the background level determined from a source free region.
The extended blue areas at the 3 detector corners are $^{55}$Fe calibration sources.
Green circles with numbers and a black circle at the center (V1647~Ori)
are X-ray source positions detected in the \CHANDRA\ observation on 2002 November 14 
\citep{Simon2004,Kastner2004}.
Radii of these circles are proportional to \CHANDRA\ photon count rates between 0.5$-$10~keV
in logarithmic scale, as shown at the bottom right corner.
The grey dots show the \CHANDRA\ ACIS-S FOV.
{\it Right}: monochromatic XIS~0+1+3 unsmoothed image of the same field between 0.5$-$10 keV.
The cyan and red circles show source and background regions, respectively; the region
within the magenta circle was excluded from these source and background regions.
\label{fig:xisimg}
}
\end{figure}

Figure~\ref{fig:xisimg} shows an XIS color image of the V1647 field, combining all the XIS detectors (0+1+3).
The XISs detected three bright peaks at the center, top, and bottom of the image
and four faint sources at $\sim$1.5\ARCMIN\ north-east, 4\ARCMIN\ south-east,
4\ARCMIN\ south, and 8\ARCMIN\ south-west from the central source.
The bright sources to the north and south and the faint source to the south-east each
have only one corresponding X-ray source in the 2002 \CHANDRA\ observation, respectively
\citep[source 14, source 8, and source 23 in][]{Simon2004}
within the \Suzaku\ pointing uncertainty 
($\lesssim$20\ARCSEC)\footnote{suzakumemo-2007-04 --- http://www.astro.isas.ac.jp/suzaku/doc/}.
We measured their source positions from a 0.5$-$10 keV XIS image smoothed with the Gaussian function
with $\sigma$=2~image pixel (16.7\ARCSEC) 
by weighting photon counts within a 3$\times$3 pixel region centered at the peak pixel.
These positions were offset, on average, by ($\Delta$R.A., $\Delta$Dec.) = ($-$8.3\ARCSEC, $-$2.2\ARCSEC)
from the coordinates in \citet{Simon2004}, with a standard deviation of 2.0\ARCSEC.
We shifted the \SUZAKU\ image frame to match the \CHANDRA\ coordinates.

After these corrections, the absolute coordinates of the central bright source 
($\alpha_{2000}, \delta_{2000}$) = (5$^{h}$46$^{m}$13.35$^{s}$, $-$00$^{d}$06\ARCMIN5.7\ARCSEC) 
correspond to those of V1647~Ori within the 3$\sigma$ error range.
Faint \CHANDRA\ sources around V1647~Ori (9, 13, 15) are on the outskirts of the PSF and do not show
any apparent local peaks at their locations in the \SUZAKU\ image.
In the \CHANDRA\ observation in September 2008 (ObsID: 9915),
these sources were far weaker than (had $\lesssim$3\% of the photon counts of) V1647 Ori.
Their contamination of V1647~Ori in the \SUZAKU\ data would therefore appear to be minimal.
The soft source to the north-east from V1647~Ori is probably a blend of sources 20 and 21 in \citet{Simon2004},
while the very hard source to the south is a blend of sources~7, 10, and 11, which are members of SSV~63, a cluster of
class~I protostars.

\section{Timing and Spectral Analysis}

\begin{figure}[t]
\plotone{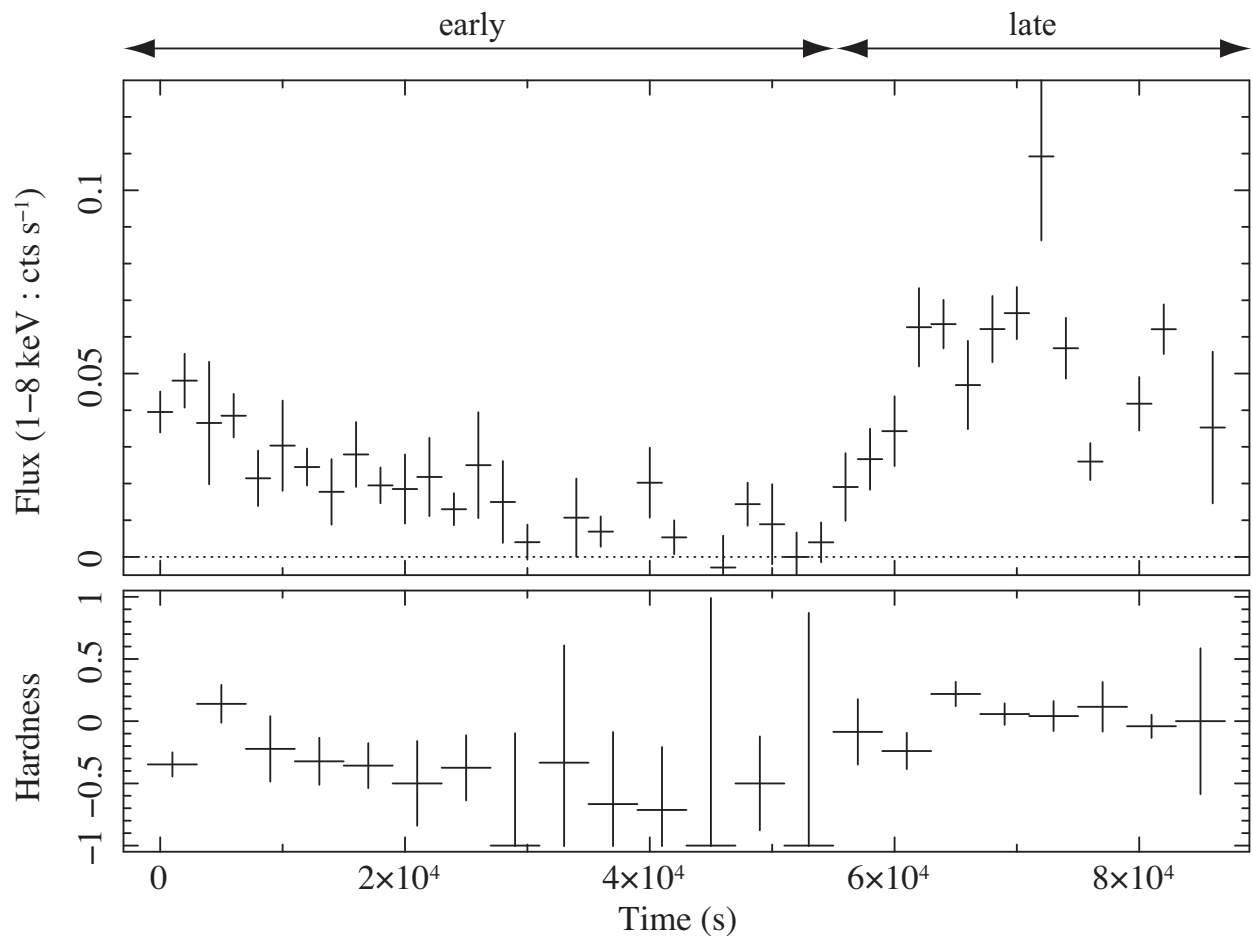}
\caption{
{\it Top panel}: Background subtracted light curve of V1647 Ori between 1$-$8~keV,
produced from the XIS~0+1+3 data. Each bin has 2~ksec.
{\it Bottom panel}: The hardness ratio curve with 4~ksec bins defined by $(H-S)/(H+S)$, 
where $H$ and $S$ are the hard (4$-$8~keV) and soft (1$-$4~keV) band count rates, respectively.
The label above the top panel shows intervals defined for phase-resolved spectral analysis.
The origin of the x-axis is 2008 October 8, 14$^{h}$33$^{m}$10$^{s}$ in UT.
\label{fig:xislcwhole}
}
\end{figure}

\begin{figure}[t]
\plotone{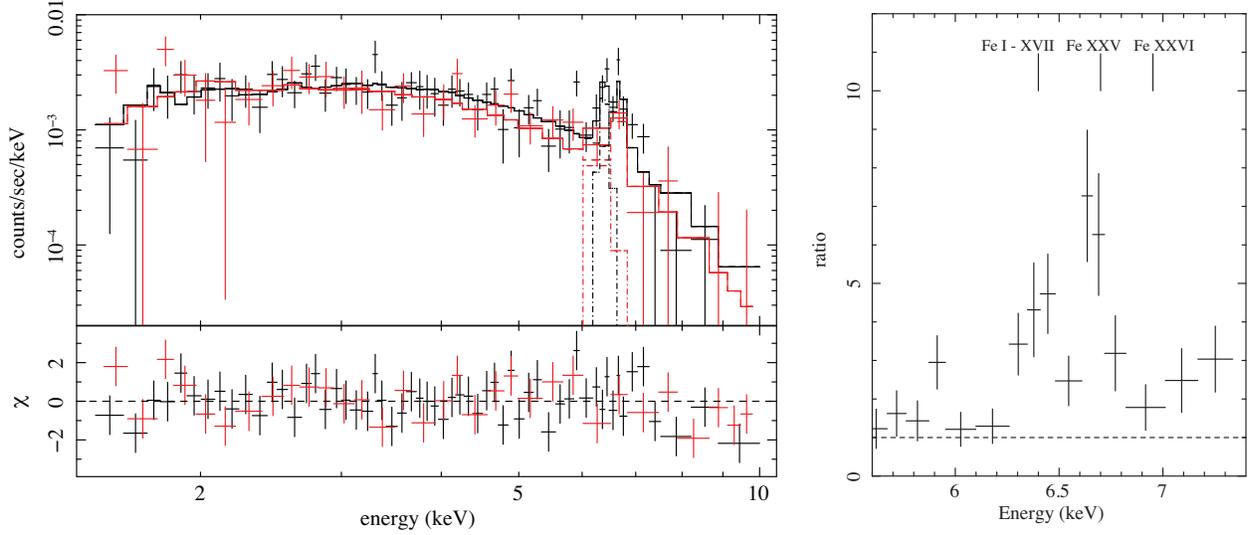}
\caption{
{\it Left Top}: XIS FI ({\it black}) and BI ({\it red}) spectra of V1647~Ori.
The solid lines show the best-fit result of the spectra by a model of 1-temperature thin-thermal
plasma emission (APEC) with a Gaussian function at 6.4~keV for the iron fluorescent line,
suffering photoelectric absorption by quasi-neutral gas.
{\it Left Bottom}: residuals of the $\chi^{2}$ values from the best-fit model.
{\it Right}: XIS 0+1+3 spectrum near the iron $K$ line region.
The vertical axis shows the ratio of the emission against the continuum in the best-fit model
(model ``i" in Table~\ref{tbl:specfit}).
Important emission lines are marked by lines at their rest energies.
\label{fig:xisspec}
}
\end{figure}

\begin{figure}[t]
\plottwo{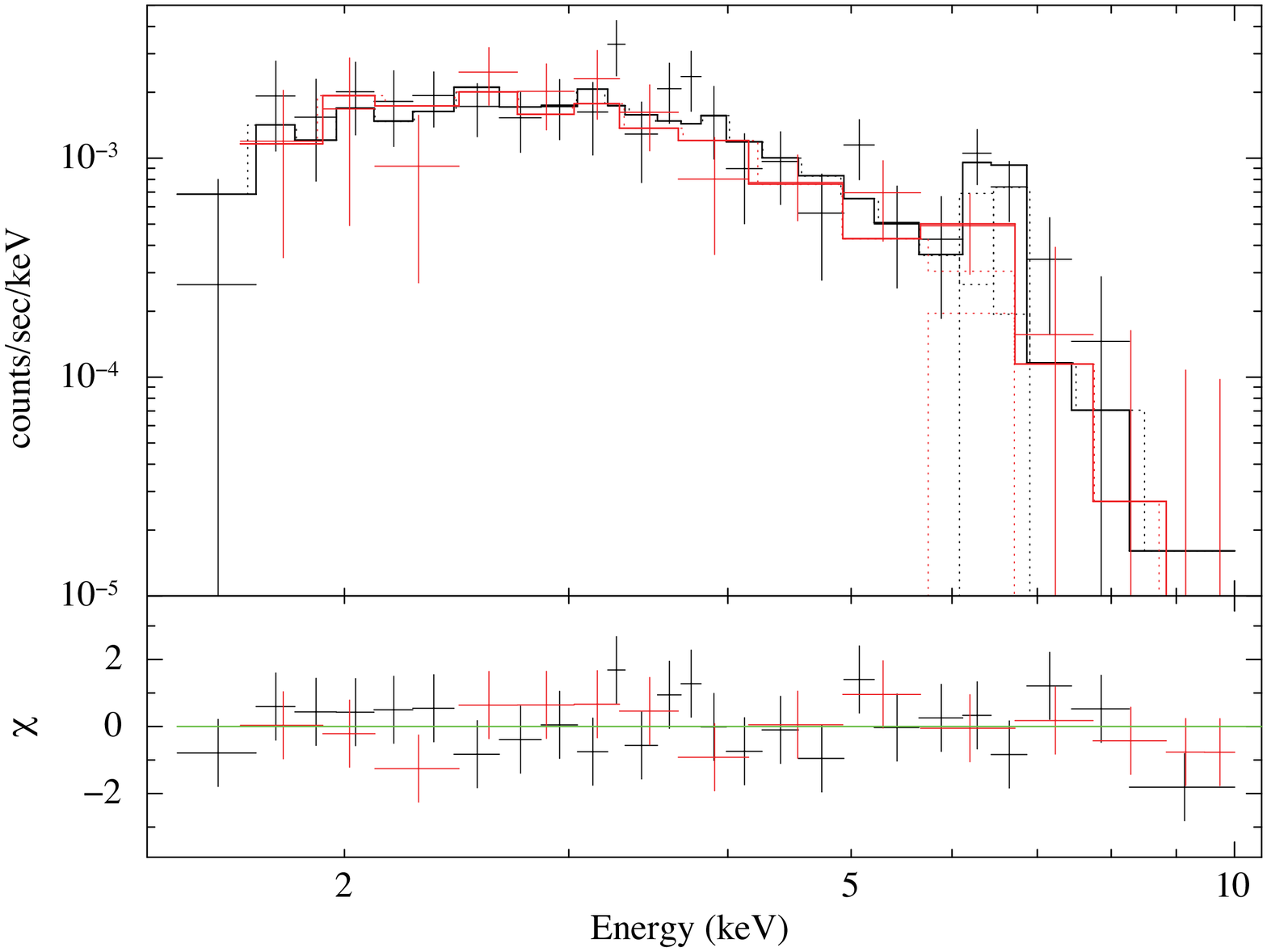}{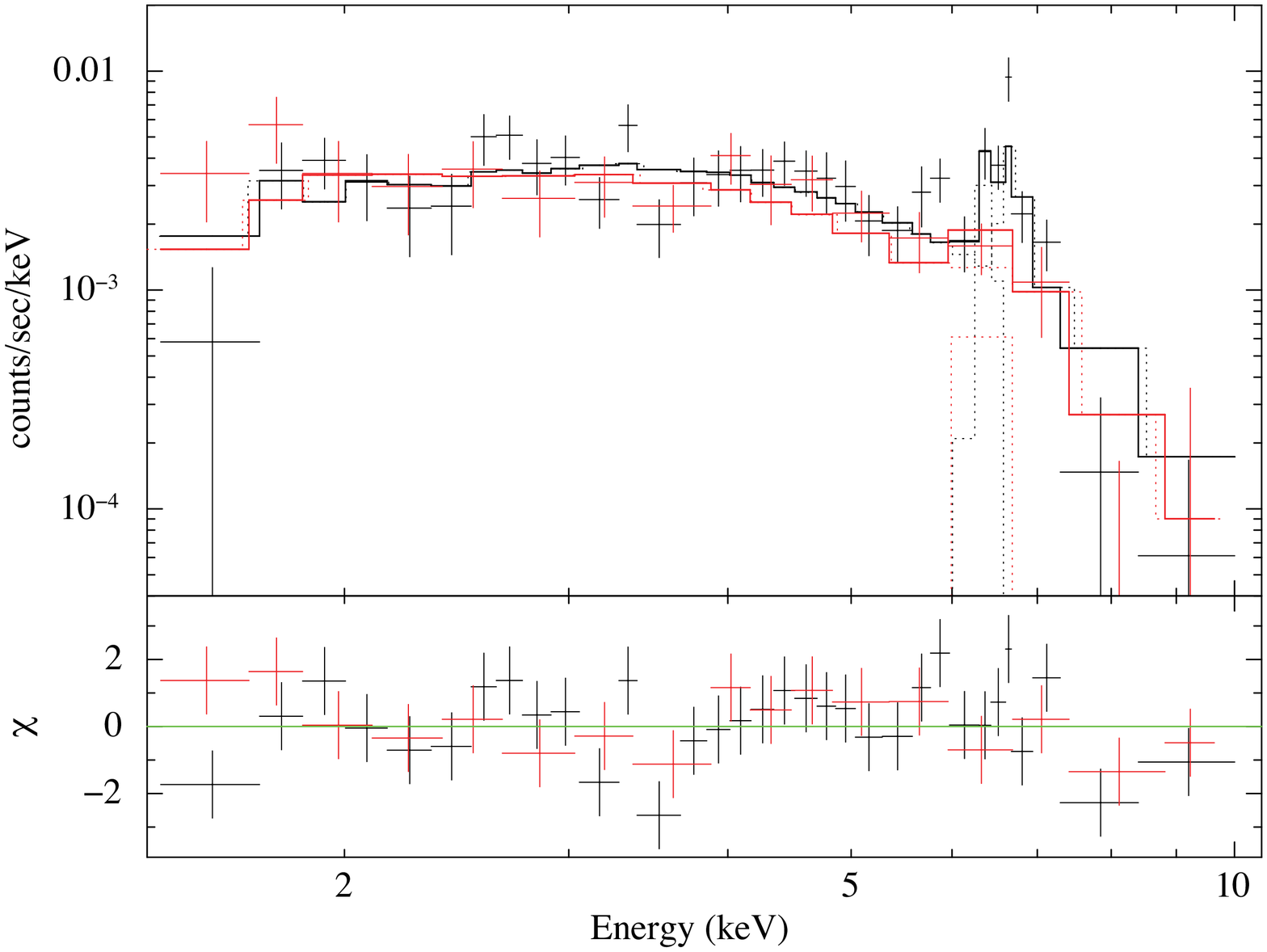}
\caption{
X-ray spectra of the early ({\it left}) and late ({\it right}) phases (model ``ii").
\label{fig:xispflflspec}
}
\end{figure}

The north-east source heavily contaminated the V1647~Ori region below $\sim$1~keV,
but its contamination was negligible above 4~keV.
To minimize the contamination and maximize the photon statistics, we defined a source region with a 1.5\ARCMIN\ 
radius circle, excluding the circular region within 40\ARCSEC\ of the north-east source (see the right panel of Figure~\ref{fig:xisimg}).
To cancel out the remaining contamination, we defined a background region that is symmetric to the source region
with respect to the north-east source.
The background-subtracted light curves and spectra presented in Figure~\ref{fig:xislcwhole}$-$\ref{fig:xispflflspec} were
generated from the events within these regions.

Figure~\ref{fig:xislcwhole} shows the background subtracted X-ray light curve of V1647~Ori covering the
energy range 1$-$8~keV, combining all the XIS (0+1+3) data, and using a bin size of 2~ks.
The few missing data points are time periods where few good-time intervals ($<$1\%) were available. 
The count rate varied strongly, by a factor of $\gtrsim$18, during the observation
(count rate of the maximum bin: $\sim$0.11~\UNITCPS, 
1$\sigma$ upper limit count rate of the minimum bin: $\sim$6$\times$10$^{-3}$~\UNITCPS).
The X-ray light curve shows a gradual decrease, to a level consistent with zero flux, over the first $\sim$60~ks of the observation, 
then a sharp increase in flux at $\sim$60~ksec followed by a decay with significant spikes and dips.

We produced background subtracted soft band (1$-$4~keV) and hard band (4$-$8~keV) light curves
and calculated the hardness ratio (HR) defined by $(H-S)/(H+S)$,
where $S$ and $H$ are the soft and hard band count rates, respectively
(bottom panel of Figure~\ref{fig:xislcwhole}).
The HR was $\sim-0.5$ between 10~ksec and 60~ksec into the observation and increased to $\sim$0.0 after $\sim$60~ksec,
remaining at that level for the rest of the observation duration.
The abrupt flux increase at $\sim$60~ksec, accompanying an HR increase, may be explained by an increase in \KT\ 
(see also spectral analysis in the latter paragraphs), while 
fluctuations in flux after that, which are apparently uncorrelated with the HR,
caused by changes in the plasma emission measure.
Such flux and HR variations were seen in the \XMM\ observation in 2004, as well \citep{Grosso2005}.

We merged spectra of the FI sensors and produced time-averaged FI and BI spectra (Figure~\ref{fig:xisspec}).
All the spectra in our analysis were binned so as to obtain at least 20 photon counts per bin.
We ignored energy bins below 1.4~keV, where contamination from the north-east source exceeds the
net signal from V1647~Ori.
The resulting spectra reveal a strong fluorescent iron line at 6.4~keV from quasi-neutral iron atoms, 
as well as the helium-like iron line at 6.7~keV, which is a signature of hot plasma.
These features can be seen clearly in the Fe K$\alpha$ line region of the 
merged XIS spectra (right panel of Figure~\ref{fig:xisspec}).
The spike at 5.9~keV possibly originates from contamination of the $^{55}$Fe calibration source scattering over 
the CCD chips.

The spectrum is well fit by a model consisting of a single-component, optically thin-thermal plasma emission (APEC)
combined with a narrow Gaussian line with the center energy fixed at 6.4~keV for the iron fluorescent emission,
suffering photoelectric absorption by neutral gas (wabs).
The best-fit parameters are listed in the model ``i" in Table~\ref{tbl:specfit}.
The best-fit plasma temperature ($\sim$4.1~keV), elemental abundance ($\sim$0.50~solar),
hydrogen column density ($\sim$3.6$\times$10$^{22}$~\UNITNH),
and the average observed flux ($\sim$4.4 $\times$10$^{-13}$~\UNITFLUX\ between 0.5$-$8~keV)
are similar to those inferred for V1647~Ori during the \XMM\ observation in 2004 \citep{Grosso2005}.
In the 2.8$-$8.0 keV band, the absorption corrected X-ray luminosity during the \SUZAKU\ observation (log~\LX~$\sim$30.9~\UNITLUMI)
was twice that seen during the \XMM\ observation (log~\LX~$\sim$30.5~\UNITLUMI).
One remarkable difference, however,  is the firm detection of the iron fluorescent line at 6.4~keV.
The equivalent width (EW) in 2008, $\sim$600~eV, was about a factor of 6 higher than that found from the marginal 
detection of the iron fluorescent line during the \XMM\ observation in 2004.

Because the spectrum changed so dramatically near the midpoint of the Suzaku exposure,
we divided the observation into two phases, before and after 56~ksec from the observation start,
as shown at the top of Figure~\ref{fig:xislcwhole}.
We simultaneously fit these spectra with their elemental abundances tied.
The best-fit results are shown in the model ``ii" in Table~\ref{tbl:specfit} and Figure~\ref{fig:xispflflspec}.
The plasma temperature was significantly lower in the early phase, while the other parameters,
especially the hydrogen column density, do not appear to vary significantly.
This would suggest that the plasma heated up and/or that a very hot plasma emerged at $\sim$56~ksec after the start of the observation.
The best-fit EW of the iron fluorescent line was huge, close to 1.5~keV, during the ``early" phase
though the error was large enough to include the best-fit EW in the ``late" phase.
The model ``iii" in Table~\ref{tbl:specfit} with their absorption and abundance parameters tied gives a similar result.
Such large fluorescent iron line EWs have been observed from spatially extended X-ray reflection sources
that do not contaminate emission from their irradiating sources \citep[e.g.,][]{Koyama1996b,Corcoran2004}.

\section{Discussion}

The \SUZAKU\ observation was performed between $\sim$40 and $\sim$280~days
after the onset of the new accretion outburst from V1647~Ori.
The first (2004) \XMM\ observation of V1647~Ori during its 2003-2005 outburst was performed with similar timing, 
$\sim$4~months after outburst onset,
with net exposure time (37$-$39~ksec) similar to that of the \SUZAKU\ observation.
Since these two observatories have similar effective areas in the 0.5$-$10~keV energy range,
these observations are well suited to comparing the X-ray properties of V1647~Ori between the 2003 and 2008 outbursts.

The flux of X-ray emission from V1647~Ori strongly varied, by at least a factor of 18, during the \SUZAKU\ observation.
The flux bottom is not well constrained given the contamination from the north-east soft source,
but the \XMM\ observation in 2004 showed a similar range of variation of between 0.5$-$12$\times$10$^{-13}$~\UNITFLUX.
Both observations showed similar abrupt flux increases with correspondingly HR increases and no significant HR variations after 
the increases.
The plasma parameters derived from the \SUZAKU\ spectra were similar to the best-fit result of the \XMM\ spectrum
\citep[model~1 in Table~\ref{tbl:specfit},][]{Grosso2005}.
These similarities suggest that the new outburst in 2008 was driven by a mechanism similar to that of the outburst
which started in 2003 October and lasted for $\sim$2~years.
Similarities between the 2003$-$2005 and latest outbursts are also seen at near-infrared wavelengths \citep{Aspin2009}.

While V1647~Ori displayed strikingly similar X-ray activity patterns during its two mass accretion outbursts,
there appears to be a very wide range of X-ray behavior and pre/post-outburst X-ray luminosities among the other FUor and EXor systems
observed in X-rays at the onset of and/or during eruptions.
The FUor systems FU Ori and V1735~Ori emit strong X-rays at levels of log~\LX~$\sim$31 \UNITLUMI\ 
from hot plasma, \KT~$>$ 5~keV \citep{Skinner2006,Skinner2009}, similar to that seen from V1647 Ori.
On the other hand, several FUor and EXor systems did not show remarkable X-ray activity associated with their mass accretion outbursts
(EXor --- V1118~Ori: \citealt{Audard2009}, FUor --- V1057~Cyg and V1515~Cyg: \citealt{Skinner2009}).
These particular X-ray characteristics are apparently not directly related to the optical outburst criteria that are used to classify
objects as FUors or EXors (e.g., outburst duration).
The X-ray behavior during outburst therefore might be more reasonably connected to the other YSO characteristics,
such as rotational velocity and/or strength of the magnetic field.

The \SUZAKU\ spectra in 2008 showed an impressively strong fluorescent iron K$\alpha$ line feature at 6.4~keV.
Such a large EW value has been reported only from a few young stellar objects
(X$_E$ or IRS~7B: \citealt{Hamaguchi2005b}, IRS1: \citealt{Skinner2007}, V1486~Ori: \citealt{Czesla2007}).
The observed iron fluorescent EWs are more than 10 times as large as that for a 4 keV plasma 
irradiating an infinite plane of solar iron abundance in a Monte-Carlo simulation \citep[$\lesssim$60~eV,][]{Drake2008},
so that special physical conditions would be required to explain such a large EW.
Plausible conditions that can elevate the EWs are
(1): supersolar iron abundance,
(2): re-emission by a neutral absorber, e.g., a flaring disk and/or infalling envelope,
(3): time-delay effect of fluorescence intensified by radiation from a magnetic flare,
(4): line-of-sight obscuration of the irradiating plasma,
(5): excitation by non-thermal electrons produced by a magnetic flare \citep[e.g.,][]{Osten2007,Czesla2007}.
Among them, the conditions (3) and (5) would require a big flare to produce the large EW,
and which was not observed during the \SUZAKU\ observation.
The condition (2) can only increase the EW by a factor of 2.
The condition (1) is unlikely because, in this condition, the photosphere and/or disk abundances would have had to increase
by a factor of six between 2004 and 2008.
We thus favor the condition (4) as the most likely explanation.

Assuming the presence of a reflector with solar iron abundance on an infinite plane, $\sim$90\% of the plasma has to be 
blocked from our direct view by an optically thick absorber in order to explain the large iron fluorescent line EW measured 
for V1647 Ori in 2008 by \SUZAKU.
Such a geometry is plausible if most of the plasma is hidden behind the stellar core and/or disk
\citep[see the discussion in][]{Hamaguchi2005b}.
We note that
\citet{Aspin2009} demonstrated a change in the $r'$ band morphology of the reflection nebula 
associated with V1647~Ori (McNeil's Nebula)
between 2004 and 2008,
and suggested that dust obscuration close to V1647~Ori may play an important role in defining the observed
morphology of the nebula.
The circumstellar gas structure very close to the stellar core may have changed between 2004 and 2008,
and this change may have altered both the reflection paths of X-ray emission and the openings in visible light between
the star and McNeil's nebula.

\acknowledgments

We greatly appreciate the Suzaku review committeeÕs positive response to our DDT request. 
This work is performed while K.H. was supported by the NASA's Astrobiology Institute (RTOP 344-53-51) 
to the Goddard Center for Astrobiology.
J.K.'s research on X-rays from erupting YSOs is supported by NASA/GSFC \XMM\ Guest Observer grant
NNX09AC11G to RIT.
% This research has made use of data obtained from the High Energy Astrophysics Science Archive
% Research Center (HEASARC), provided by NASA's Goddard Space Flight Center.

Facilities: \facility{Suzaku (XIS)}

\bibliographystyle{apj}

\begin{deluxetable}{ccccccccc}
\tablecolumns{9}
\tablewidth{0pc}
\tabletypesize{\scriptsize}
\tablecaption{Spectral Fits\label{tbl:specfit}}
\tablehead{
\colhead{Model}&
\colhead{Phase}&
\colhead{\KT}&
\colhead{$Z$}&
\colhead{\NH}&
\colhead{Line Flux$^{a}$}&
\colhead{Line EW$^{b}$}&
\colhead{log \LX$^{c}$}&
\colhead{$\chi^{2}$~($d.o.f.$)}
\\
&&\colhead{(keV)}&\colhead{(solar)}&\colhead{(10$^{22}$ \UNITNH)}&\colhead{(10$^{-6}$ ph cm$^{-2}$ s$^{-1}$)}&
\colhead{(eV)}&\colhead{(\UNITLUMI)}
}
\startdata
i&whole&4.1~(3.2$-$5.6)&0.50~(0.31$-$0.73)&3.6~(2.9$-$4.4)&2.7~(1.6$-$3.8)&588&31.1/30.9/31.3&1.03~(85)\\
ii&early&1.9~(1.4$-$2.7)&0.65~(0.36$-$1.02)&5.1~(3.7$-$7.1)&2.1~(0.66$-$3.6)&1328&31.3/30.7/31.4&1.00~(84)\\
&late&5.3~(3.8$-$7.5)&=early&3.6~(2.9$-$4.5)&4.5~(2.3$-$6.8)&549&31.2/31.2/31.5&\nodata\\
iii&early&2.3~(1.8$-$3.2)&0.62~(0.32$-$0.97)&4.0~(3.3$-$4.9)&2.4~(0.33$-$3.5)&1355&31.1/30.7/31.2&1.02~(85)\\
&late&4.7~(3.5$-$6.8)&=early&=early&3.8~(2.1$-$7.0)&459&31.2/31.2/31.5&\nodata\\ \hline
1$^{d}$&2004 April 4&3.0~(2.4$-$3.9)&0.8~(0.5$-$1.3)&2.9~(2.5$-$3.4)&\nodata&\nodata&30.8/30.5/31.0&1.11~(129)\\
\enddata
\tablecomments{
$^{a}$Total photon flux in the narrow Gaussian line with a fixed center at 6.4 keV.
$^{b}$Measured with the xspec12 command ``eqw range 0 {\it Gaussian\_component\_ID}".
$^{c}$Absorption corrected luminosity between 0.5$-$2.8/2.8$-$8/0.5$-$8~keV, assuming $d =$400~pc.
$^{d}$Model \#1 in Table 1 of \citet{Grosso2005}.
The parentheses between the second and fifth columns show 90\% confidence ranges.
}
\end{deluxetable}

\clearpage

\end{document}